%% file: wfjump.tex
\documentclass[onecolumn,prd,aps,superscriptaddress,nofootinbib]{revtex4}

\usepackage{bm}
\usepackage{latexsym}
\usepackage{amsmath}
\usepackage{bm}
\usepackage{graphicx}
\usepackage{hyperref}
\usepackage{xspace}
\usepackage{amsfonts}

\newcommand{\mapPlanck}{\includegraphics[width=0.7cm,height=0.3cm,trim=-1cm 2cm 0
0]{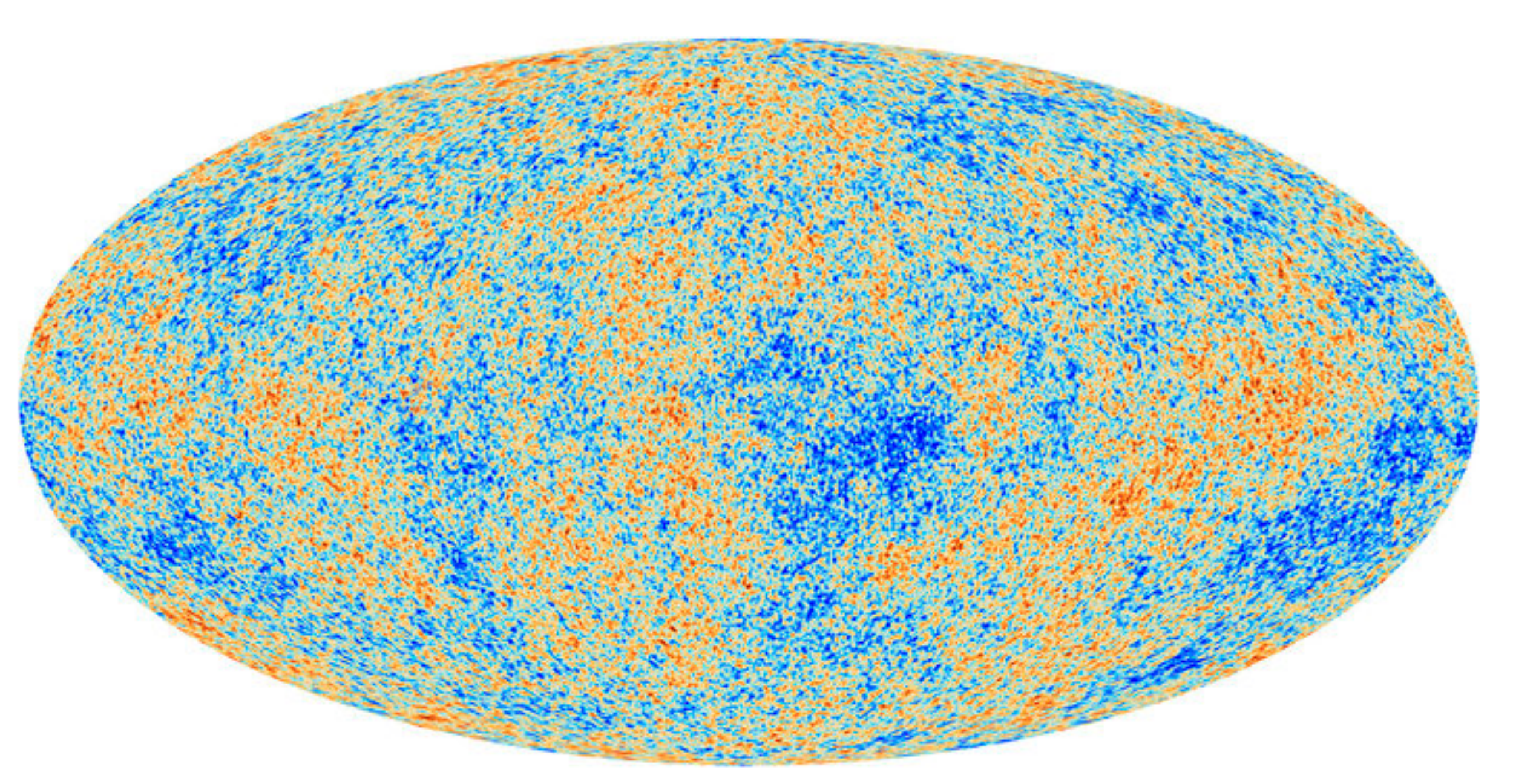}}
\newcommand{\ketmapPlanck}{\vert\mapPlanck\rangle}

\newcommand{\map}{\includegraphics[width=0.7cm,height=0.3cm,trim=-1cm 2cm 0
0]{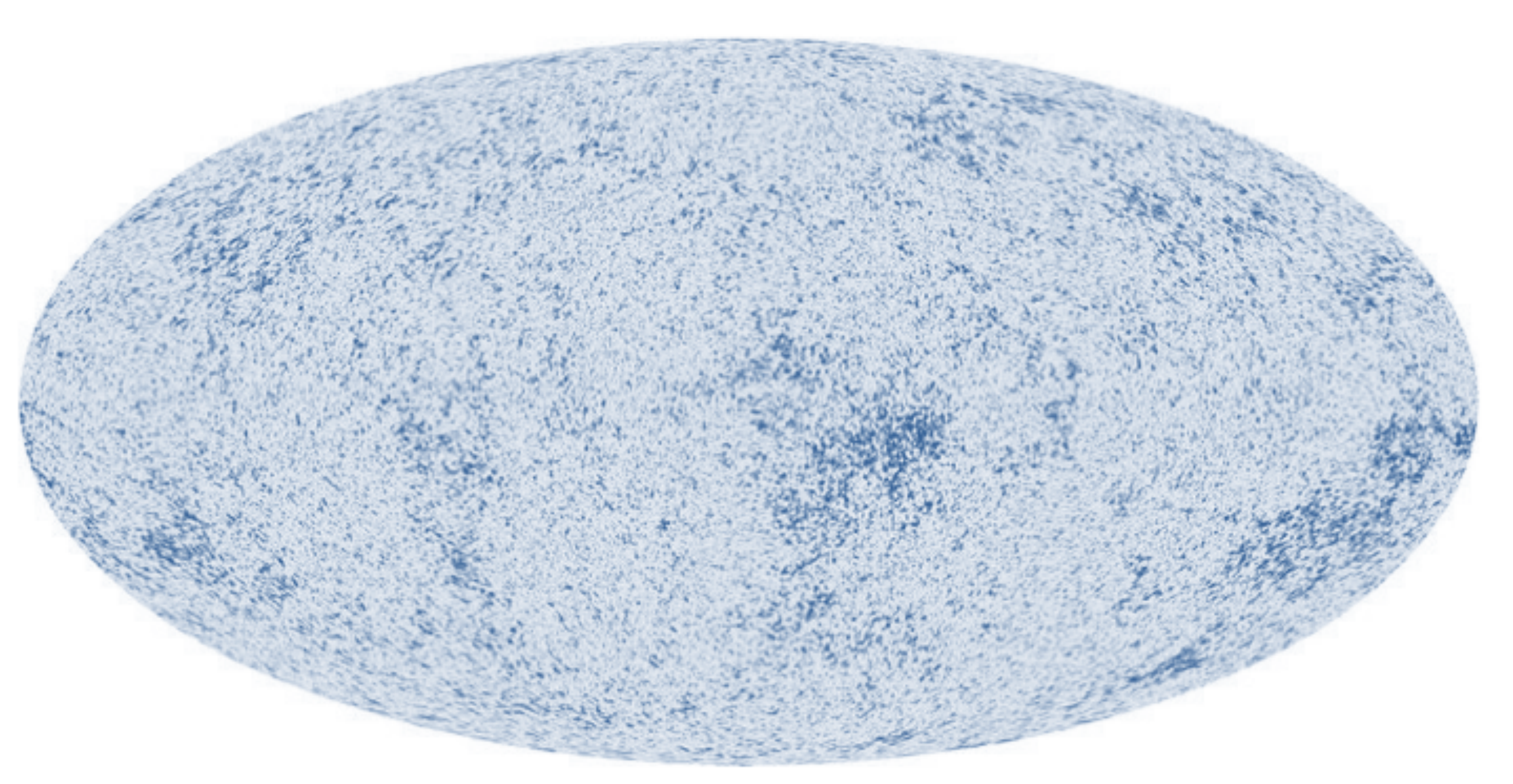}}
\newcommand{\ketmap}{\vert\map\rangle}

\makeatletter
\gdef\@fpheader{}
\g@addto@macro\bfseries{\boldmath}
\makeatother

\hypersetup{
    bookmarks=true,         % show bookmarks bar?
    unicode=false,          % non-Latin characters in Acrobat’s bookmarks
    pdftoolbar=true,        % show Acrobat’s toolbar?
    pdfmenubar=true,        % show Acrobat’s menu?
    pdffitwindow=false,     % window fit to page when opened
    pdfstartview={FitH},    % fits the width of the page to the window   
    pdfnewwindow=true,      % links in new window
    colorlinks=true,        % false: boxed links; true: colored links
    linkcolor=red,          % color of internal links
    citecolor=cyan,         % color of links to bibliography
    filecolor=magenta,      % color of file links
    urlcolor=blue,         % color of external links
    linktocpage=true
}

\input{newcommands}

\begin{document}

\title{Collapse Models and Cosmology}

\author{J\'er\^ome Martin} \email{jmartin@iap.fr}
\affiliation{Institut d'Astrophysique de Paris, UMR 7095-CNRS,
  Universit\'e Pierre et Marie Curie, 98bis boulevard Arago, 75014
  Paris, France}

\author{Vincent Vennin} \email{vincent.vennin@apc.in2p3.fr}
\affiliation{Laboratoire Astroparticule et Cosmologie, Universit\'e
  Denis Diderot Paris 7, 75013 Paris, France} \affiliation{Institut
  d'Astrophysique de Paris, UMR 7095-CNRS, Universit\'e Pierre et
  Marie Curie, 98bis boulevard Arago, 75014 Paris, France}

\date{\today}

\begin{abstract}
  Attempts to apply quantum collapse theories to Cosmology and cosmic
  inflation are reviewed. These attempts are motivated by the fact
  that the theory of cosmological perturbations of quantum-mechanical
  origin suffers from the single outcome problem, which is a modern
  incarnation of the quantum measurement problem, and that collapse
  models can provide a solution to these issues. Since inflationary
  predictions can be very accurately tested by cosmological data, this
  also leads to constraints on collapse models. These constraints are
  derived in the case of Continuous Spontaneous Localization (CSL) and
  are shown to be of unprecedented efficiency.
\end{abstract}

\maketitle

\section{Introduction}
\label{sec:introduction}

Quantum Mechanics finds itself in a somehow paradoxical situation. On
one hand, it is an extremely efficient and well-tested theory whose
experimental successes are impressive and unquestioned. On the other
hand, understanding and interpreting the formalism on which it rests
is still a matter of debates. This on-going discussion has led to a
variety of points of view ranging from challenging that there is an
actual problem, to developing different ways of understanding the
theory or, in other words, different
``interpretations''~\cite{Bassi:2012bg}.

Giancarlo Ghirardi, to whom this book and chapter are dedicated, has
made fundamental contributions to this question. In fact, the approach
proposed by Ghirardi (together with his collaborators, Rimini and
Weber and, independently, Pearle), the so-called collapse
models~\cite{Ghirardi:1985mt,Pearle:1988uh,Ghirardi:1989cn,
  Bassi:2003gd}, unlike the other interpretations, goes beyond simply
advocating for a different scheme to capture the meaning of the
Quantum Mechanics formalism. It is actually an alternative to Quantum
Mechanics and, as such, it should not be considered as an
interpretation but rather as another, rival, theory. In some sense,
collapse models enlarge Quantum Mechanics, which becomes only one
particular theory in a larger parameter space, in the same way that,
for instance, General Relativity is only one point in the parameter
space of scalar-tensor theories~\cite{EspositoFarese:2004cc}. As a
consequence, the great advantage of collapse theories is that they
make predictions that are different from those of Quantum Mechanics
and that can thus be falsified. This was of course realized from the
very beginning by Ghirardi and, nowadays, there exists a long list of
experiments aiming at constraining collapse
models~\cite{Bassi:2012bg}.

These experiments, however, are all performed in the lab. In the present
article, it is pointed out that using Quantum Mechanics and/or
collapse models in a cosmological context can shed new light on those theories.

One of the most important insights in Cosmology is the realization that
galaxies are of quantum-mechanical
origin~\cite{Mukhanov:1981xt}. They are indeed nothing but quantum
fluctuations, stretched to very large distances by cosmic expansion
during a phase of
inflation~\cite{Starobinsky:1980te,Guth:1980zm,Linde:1981mu,Albrecht:1982wi,Linde:1983gd}
and amplified by gravitational instability. This discovery has clearly
far-reaching implications for Cosmology but also for foundational
issues in Quantum Mechanics. Indeed, in Cosmology, Quantum Mechanics
is pushed to new territories not only in terms of scales (the typical
energy, length or time scales relevant for Cosmology are very
different from those characterizing lab experiments) but also in terms
of concepts: applying Quantum Mechanics to a single system with no
exterior, classical, domain is not trivial~\cite{1955mfqm.book.....V, Hartle:2019hae}.

Among the first physicists who realized that Cosmology can be an
interesting playground for Quantum Mechanics was John~Bell, see for
instance his article ``{\it Quantum mechanics for
  cosmologists}''~\cite{1988PhT....41j..89B}. As Ghirardi recalled and
discussed in detail during the colloquium he gave at the Institut
d'Astrophysique de Paris (IAP) on March 22nd, 2012, he and John Bell
were good friends and enjoyed interacting together. In his
talk,\footnote{The slides of his talk can be found at this
  \href{http://www.iap.fr/vie_scientifique/seminaires/Seminaire_GReCO/2012/presentations/ghirardi.pdf}{URL}.}
Ghirardi mentioned that Bell emphasized the importance of developing a
relativistic, Lorentz invariant, version of collapse models which is
of course a prerequisite for Cosmology. He also stressed that one
important feature of collapse models is that there is ``no mention of
measurements, observers and so on'', a property that is clearly
relevant for Cosmology. Therefore, even if Ghirardi never explicitly
worked at the interface between Cosmology and Quantum Foundations, he
clearly considered this subject as a promising direction of research.

Recently, the collapse models have started to be considered in
Cosmology~\cite{Perez:2005gh, Pearle:2007rw, Lochan:2012di,
  Martin:2012pea, Canate:2012ua, Piccirilli:2017mto, Leon:2017yna,
  Leon:2019jsl,Martin:2019jye}, in particular in the context of cosmic
inflation, with two essential motivations: to avoid conceptual
problems related to the absence of an observer in the very early
universe; and to use the high-accuracy cosmological data constraining
inflation as a probe of the free parameters characterizing collapse
models~\cite{Martin:2019jye}. The goal of this paper is to briefly
review these recent works. It is organized as follows. In the next
section, Sec.~\ref{sec:inflation}, we briefly review cosmic inflation
and the theory of cosmological perturbations of quantum-mechanical
origin. Then, in Sec.~\ref{sec:motivation}, we explain why collapse
theories can be useful in Cosmology. In Sec.~\ref{sec:infcoll}, we
discuss how these theories can be implemented concretely and, in
Sec.~\ref{sec:comparison}, we use cosmological observations to put
constraints on the parameters characterizing collapse models. Finally,
in Sec.~\ref{sec:conclusions}, we present our conclusions.

\section{Cosmic inflation and cosmological perturbations}
\label{sec:inflation}

In Cosmology, the theory of inflation is a description of the physics
of the very early
universe~\cite{Starobinsky:1980te,Guth:1980zm,Linde:1981mu,Albrecht:1982wi,Linde:1983gd}. It
is a phase of exponential, accelerated, expansion [meaning that
  $\ddot{a}>0$ where $a(t)$ is the scale factor describing how cosmic
  expansion proceeds and $t$ is the cosmic time] first introduced to
fix some undesirable features of the standard model of
Cosmology~\cite{Martin:2019zia}. Since it occurs in the
early universe, it is characterized by a very high energy scale, that
could be as large as $10^{15}\GeV$. Soon after inflation was proposed,
in the late seventies and early eighties, it was also realized that it
provides an efficient mechanism for structure formation. In the
present context, ``structures'' refer to the small inhomogeneities
that are the seeds of the Cosmic Microwave Background (CMB)
anisotropies and of the galaxies. They can be represented by an
inhomogeneous scalar field called the ``curvature
perturbation''~\cite{Mukhanov:1981xt,Kodama:1985bj}, and denoted
$\zeta(t,\bm{x})$. It represents small ripples propagating on top of
an homogeneous and isotropic background. The idea is then to promote
this scalar field to a quantum scalar field, which thus undergoes
unavoidable quantum fluctuations. These quantum fluctuations are then
amplified during inflation and, later on in the history of the
universe, give rise to galaxies.

This may seem a rather drastic idea, but one can show that all the
predictions of this theory are in perfect agreement with astrophysical
observations~\cite{{Ade:2013zuv,Martin:2013tda,Martin:2013nzq,Martin:2014nya,Martin:2015dha,Akrami:2018vks,
    Akrami:2018odb}}. In particular, the statistics of $\zeta$ are
quasi Gaussian (no deviation from Gaussianity has been detected so
far~\cite{Akrami:2019izv}), and can thus be fully characterized in
terms of its power spectrum ${\cal P}_\zeta(k)$, which is the square
of its Fourier amplitude. It represents the ``amount'' of
inhomogeneities at a given scale. It was known as an empirical fact,
well before the advent of inflation, that cosmological data are
consistent with a primordial scale-invariant power spectrum, that is
to say with a function ${\cal P}_\zeta(k)$ that is
$k$-independent. But the theoretical origin of this scale-invariance
was not known. Inflation definitively gained respectability when it
was realized that it leads to this type of power spectrum for the
quantum fluctuations mentioned before. Its convincing power is even
higher today because, in fact, inflation does not predict an exact
scale-invariant power spectrum, but rather an almost scale-invariant
power spectrum: if one writes the power spectrum as ${\cal
  P}_\zeta(k)\sim k^{\nS-1}$, where $\nS$ is the so-called spectral
index, exact scale-invariance corresponds to $\nS=1$ while inflation
leads to $\nS\neq 1 $ but $\vert\nS-1\vert \ll 1$. As a consequence,
if inflation is correct, then one should observe a small deviation
from $\nS=1$. In $2013$, the European Space Agency (ESA) satellite
Planck measured the CMB anisotropies with exquisite precision and
found~\cite{Ade:2013zuv} $\nS=0.9603\pm 0.0073$, thus establishing
that, if $\nS$ is indeed close to one, it differs from one at a
(5$\sigma$) significant level. The most recent
release~\cite{Akrami:2018vks, Akrami:2018odb}, in $2018$, has
confirmed this measurement with $\nS=0.9649\pm 0.0042$. This
confirmation of a crucial inflationary prediction has given a strong
support to the idea that galaxies are of quantum-mechanical origin.

At the technical level, it is well known that a field in flat
space-time can be interpreted as an infinite collection of harmonic
oscillators, each oscillator corresponding to a given Fourier
mode. Likewise, a scalar field living in a cosmological, curved,
space-time can be viewed as an infinite collection of
\emph{parametric} oscillators, the fundamental frequency of each
oscillator becoming a time-dependent function because of cosmic
expansion (for a review, see \Refa{Martin:2007bw}). Upon quantization,
harmonic oscillators naturally lead to the concept of coherent states
while parametric oscillators lead to the concept of squeezed
states~\cite{CohenTannoudji:1992}. In the Heisenberg picture, the
curvature perturbation operator can be expanded as
\begin{align}
  \label{eq:operzeta}
  \hat{\zeta}(\eta,{\bm x})=\frac{1}{(2\pi)^{3/2}}
  \frac{1}{z(\eta)}\int \frac{{\rm d}{\bm k}}{\sqrt{2k}}\left[\hat{c}_{\bm k}(\eta)
  e^{i{\bm k}\cdot {\bm x}}+\hat{c}_{\bm k}^{\dagger}(\eta)
  e^{-i{\bm k}\cdot {\bm x}}\right],
\end{align}
where $\hat{c}_{\bm k}(\eta)$ and $\hat{c}_{\bm k}^{\dagger}(\eta)$
are the annihilation and creation operators satisfying the usual
equal-time commutation relations, $[\hat{c}_{\bm k}(\eta),\hat{c}_{\bm
    p}^{\dagger}(\eta)]=\delta({\bm k}-{\bm p})$, $z(\eta)$ is a
function that depends on the scale factor and its derivatives only,
and $\eta$ denotes the conformal time, related to cosmic time via $\dd
t= a \dd\eta$. The dynamics of $\hat{\zeta}(\eta,{\bm x})$ is
controlled by the following Hamiltonian, which is directly obtained
from expanding the Einstein-Hilbert action plus the action of a scalar
field at second order\footnote{This second-order expansion of the action is valid at linear order in perturbation theory, which is known to provide an excellent description of primordial fluctuations, given their small amplitude. This is the order at which the calculation is performed in this work, as in the standard treatment. At higher order, mode coupling effects are expected, which would made the use of the CSL theory technically more challenging (as for the case of standard quantum mechanics) but these effects are clearly suppressed by the amplitude of perturbations, hence they cannot change our conclusions.} in perturbation theory~\cite{Martin:2007bw},
\begin{align}
  \label{eq:hamilton}
  \hat{H}=\int _{\setR^3}{\rm d}^3{\bm k}\, \hat{H}_\mathrm{free}({\bm k})
  +g(\eta)\int _{\setR^3}{\rm d}^3{\bm k}\, H_\mathrm{int}({\bm k}).
  \end{align}
In this expression, $g(\eta)=z'/(2z)$ is a time-dependent ``coupling
constant'', and
\begin{align}
\label{eq:hamil}
  \hat{H}_\mathrm{free}({\bm k})=
  \frac{k}{2}\left(\hat{c}_{\bm k}\hat{c}_{\bm k}^{\dagger}
  +\hat{c}_{-{\bm k}}^{\dagger}\hat{c}_{-{\bm k}}\right),
  \quad
  \hat{H}_\mathrm{int}({\bm k})=-i\left(\hat{c}_{\bm k}\hat{c}_{-{\bm k}}
  -\hat{c}_{-{\bm k}}^{\dagger}c_{\bm k}^{\dagger}\right).
\end{align}
The first term, $\hat{H}_\mathrm{free}$, is the Hamiltonian of a
collection of harmonic oscillators and the second one,
$\hat{H}_\mathrm{int}$, represents the interaction of the quantum
perturbations with the classical background. If space-time is not
dynamical (Minkowski), then $g(\eta)=0$. In the inflationary paradigm,
a crucial assumption, without which the theory would not be
empirically successful, is that the initial state of the system is the
so-called ``Bunch-Davies'' or ``adiabatic'' vacuum
state~\cite{Bunch:1978yq}, which can be written as
\begin{align}
  \vert 0\rangle =\bigotimes _{\bm k}\vert 0_{\bm k}\rangle,
  \end{align}
with $\hat{c}_{\bm k}(\eta_\uini)\vert 0_{\bm k}\rangle =0$,
$\eta_\uini$ being the conformal time at which the initial state is
chosen. The time evolution of the curvature perturbation
$\hat{\zeta}(\eta,{\bm x})$ is then given by the Heisenberg equation
$\dd \hat{c}_{\bm k}/\dd \eta=-i[\hat{c}_{\bm k},\hat{H}]$. This
equation can be solved by means of a Bogoliubov transformation,
$\hat{c}_{\bm k}(\eta)=u_k(\eta)\hat{c}_{\bm k}(\eta_\uini)+v_k(\eta)
\hat{c}_{-{\bm k}}^{\dagger}(\eta_\uini)$, where the functions $u_{
  k}(\eta)$ and $v_k(\eta)$ obey
\begin{align}
  \label{eq:equv}
  i\frac{\dd u_k}{\dd \eta}=ku_k(\eta)+i\frac{z'}{z}v_k^*(\eta),
  \quad 
  i\frac{\dd v_k}{\dd \eta}=kv_k(\eta)+i\frac{z'}{z}u_k^*(\eta).
  \end{align}
These functions must satisfy $\vert u_k(\eta)\vert^2-\vert
v_k(\eta)\vert^2=1$ in order for the commutation relation between
$\hat{c}_{\bm k}$ and $\hat{c}_{\bm p}^{\dagger}$ to be satisfied. If
one introduces the Bogoliubov transformation into the
expression~(\ref{eq:operzeta}) for the curvature operator, one obtains
\begin{align}
  \label{eq:operzetaini}
  \hat{\zeta}(\eta,{\bm x})=\frac{1}{(2\pi)^{3/2}}
  \frac{1}{z(\eta)}\int \frac{{\rm d}{\bm k}}{\sqrt{2k}}
  \left[(u_k+v_k^*)(\eta)\hat{c}_{\bm k}(\eta_\uini)
  e^{i{\bm k}\cdot {\bm x}}+(u_k^*+v_k)(\eta)\hat{c}_{\bm k}^{\dagger}(\eta_\uini)
  e^{-i{\bm k}\cdot {\bm x}}\right].
\end{align}
From \Eqs{eq:equv}, it is easy to establish that the quantity
$u_k+v_k^*$ obeys the equation $(u_k+v_k^*)''+\omega^2(u_k+v_k^*)=0$
with $\omega^2=k^2-z''/z$. This is the equation of a parametric
oscillator, namely a harmonic oscillator with time-dependent
fundamental frequency, and, here, this time dependence is entirely
controlled by the dynamics of the underlying background space-time. Let
us notice that the initial conditions are given by $u_k(\eta_\uini)=1$
and $v_k(\eta_\uini)=0$, which implies that
$(u_k+v_k^*)(\eta_\uini)=1$. Having solved the time evolution of the
system, one can then calculate the two-point correlation function of
the curvature perturbation. It needs to be evaluated in the state
$\vert0\rangle $ since, in the Heisenberg picture, states do not
evolve in time, and one has
\begin{align}
  \label{eq:twopointheisen}
  \left\langle 0\left \vert \zeta ^2\left(\eta,{\bm x}\right) \right
  \vert 0\right \rangle \equiv \int _0^{+\infty} \frac{\dd k}{k}{\cal
    P}_\zeta(k) =\int_0^{\infty}\frac{\dd k}{k}k^2 \left\vert
  \frac{u_k+v_k^*}{z}\right\vert^2\, .
\end{align}
This shows how the power spectrum ${\cal P}_\zeta(k)$ mentioned above
can be determined explicitly once the differential equation for
$u_k+v_k^*$ has been solved. Notice that it is, a priori, a function
of time. However, on large scales, $u_k+v_k^* \propto z$, and this
time dependence disappears.

Let us now describe the same phenomenon but in the Schr\"odinger
picture.  We first notice that the Bogoliubov transformation
introduced above can be written
\begin{align}
  \hat{c}_{\bm k}(\eta)=\hat{R}_k^{\dagger}\hat{S}_k^{\dagger}
  \hat{c}_{\bm k}(\eta_\uini)\hat{S}_k\hat{R}_k,
  \end{align}
where the operators $\hat{R}_k$ and $\hat{S}_k$, called the rotation
and squeezing operators respectively, are defined by
$\hat{R}_k=e^{\hat{D}_k}$ and $\hat{S}_k=e^{\hat{B}_k}$, with
\begin{align}
  \hat{B}_{\bm k}=r_ke^{-2i\varphi_k}\hat{c}_{-{\bm k}}(\eta_\uini)
  \hat{c}_{\bm k}(\eta_\uini)-
r_ke^{2i\varphi_k}\hat{c}_{-{\bm k}}^{\dagger}(\eta_\uini)
  \hat{c}_{\bm k}^{\dagger}(\eta_\uini), \quad 
\hat{D}_{\bm k}=-i\theta_{k,1}\hat{c}_{{\bm k}}^{\dagger}(\eta_\uini)
  \hat{c}_{\bm k}(\eta_\uini)-
i\theta_{k,2}\hat{c}_{-{\bm k}}^{\dagger}(\eta_\uini)
  \hat{c}_{-{\bm k}}(\eta_\uini).
\end{align}
They are expressed in terms of the squeezing parameter $r_k(\eta)$,
the squeezing angle $\varphi_k(\eta)$ and the rotation angle
$\theta_{k}(\eta)\equiv \theta_{k,1}(\eta)=\theta_{k,2}(\eta)$, which
are related to the functions $u_{ k}(\eta)$ and $v_k(\eta)$ via $u_{
  k}(\eta) = \ee^{-i \theta_k} \cosh r_k$ and $v_k(\eta) = -i \ee^{i
  \theta_k+2i \varphi_k}\sinh r_k$. In the Schr\"odinger picture, the
state evolves with time into a two-mode squeezed
state~\cite{Grishchuk:1990bj}
\begin{align}
\label{eq:quantumstate}
\vert 0\rangle \rightarrow
\vert \Psi_{2 {\rm sq}}\rangle=
\bigotimes_{\bm k}\hat{S}_k\hat{R}_k
\vert 0_{\bm k},0_{-{\bm k}}\rangle=
\bigotimes_{\bm k} \frac{1}{\cosh r_k(\eta)}
\sum_{n=0}^{\infty}e^{-2in \varphi_k(\eta)}
\tanh ^nr_k(\eta) \vert n_{\bm k},n_{-{\bm k}}\rangle,
\end{align}
where $\vert n_{\bm k}\rangle $ is an eigenvector of the particle
number operator in the mode ${\bm k}$. In Cosmology, the value of the
squeezing parameter, for the modes ${\bm k}$ probed in the CMB, is
$r_k\simeq 10^2$ towards the end of inflation, which is much larger
than what can be achieved in the lab. Moreover, this state is, as
apparent on the previous expression, entangled. It is therefore
reasonable to conclude that the quantum state $\vert \Psi_{2 {\rm
    sq}}\rangle$ is a highly non-classical state.

The above squeezed state can also be written in terms of a
wave-functional, which usually corresponds to writing the state in the
``position'' basis. This, however, is not as straightforward as it might
seem in the present context. Indeed, the curvature perturbation and its
conjugate momentum are related to the creation and annihilation
operators through
\begin{align}
\label{eq:zetacrea}
z(\eta)\hat{\zeta}_{\bm k}=\frac{1}{\sqrt{2k}}\left(\hat{c}_{\bm k}
+\hat{c}_{-{\bm k}}^{\dagger}\right), \quad 
z(\eta)\hat{\zeta}_{\bm k}'=-i\sqrt{\frac{k}{2}}\left(\hat{c}_{\bm k}
-\hat{c}_{-{\bm k}}^{\dagger}\right).
\end{align}
We notice that the curvature perturbation and its conjugate momentum are
not Hermitian operators since the above relations imply that
$\hat{\zeta}_{\bm k}^{\dagger}=\hat{\zeta}_{-{\bm k}}$, which simply translates
 the fact that the curvature perturbation is a real field. As a
consequence, $\hat{\zeta}_{\bm k}$ cannot play the role of the
position operator. Moreover, these expressions mix creation and annihilation
operators of momentum ${\bm k}$ and $-{\bm k}$, while it seems more natural to
define a position operator for each mode ${\bm k}$. This, however, can
be done if one introduces the operators $\hat{q}_{\bm k}$ and
$\hat{\pi}_{\bm k}$ defined by~\cite{Martin:2015qta}
\begin{align}
z(\eta)   \hat{\zeta}_{\bm k}=\frac12\left[\hat{q}_{\bm k}+\hat{q}_{-{\bm k}}
    +\frac{i}{k}\left(\hat{\pi}_{\bm k}-\hat{\pi}_{-{\bm k}}\right)\right], \quad 
z(\eta)  \hat{\zeta}_{\bm k}'=\frac{1}{2i}
\left[k\left(\hat{q}_{\bm k}-\hat{q}_{-{\bm k}}
    \right)+i\left(\hat{\pi}_{\bm k}+\hat{\pi}_{-{\bm k}}\right)\right].
    \end{align}
From those relations, it is easy to establish that
\begin{align}
\label{eq:qpcrea}
\hat{q}_{\bm k}=\frac{1}{\sqrt{2k}}\left(\hat{c}_{\bm k}
+\hat{c}_{{\bm k}}^{\dagger}\right), \quad 
\hat{\pi}_{\bm k}=-i\sqrt{\frac{k}{2}}\left(\hat{c}_{\bm k}
-\hat{c}_{{\bm k}}^{\dagger}\right),
\end{align}
so that $\hat{q}_{\bm k}$ and $\hat{\pi}_{\bm k}$ involve only
creation and annihilation operators for a fixed mode ${\bm k}$. It is
also easy to check that $[\hat{q}_{\bm k},\hat{\pi}_{\bm k}]=i$, such
that $\hat{q}_{\bm k}$ and $\hat{\pi}_{\bm k}$ are the proper
generalization of ``position'' and ``momentum'' for field
theory. Then, it follows that the total wave-functional of the system
can be written as a product of wave-functions for each mode, namely
$\Psi_\mathrm{2 sq}[\eta;q]=\prod _{{\bm k}}\Psi_{\bm k}(q_{\bm
  k},q_{-{\bm k}})$, with
\begin{align}
  \Psi_{{\bm k}}\left(q_{\bm k},q_{-{\bm k}}\right)
  =\left\langle q_{\bm k},q_{-{\bm k}}\vert \Psi_{{\bm k}}\right \rangle
  =\frac{e^{A\left(r_k,\varphi_k\right)(q_{\bm k}^2+q_{-{\bm k}}^2)-B\left(r_k,\varphi_k\right)q_{\bm k}q_{-{\bm k}}}}{\cosh r_k \sqrt{\pi}
    \sqrt{1-e^{-4i\varphi_k}\tanh ^2 r_k}},
  \end{align}
where the functions $A(r_k,\varphi_k)$ and $B(r_k, \varphi_k)$ are
defined by
\begin{align}
  A\left(r_k,\varphi_k\right)=\frac{e^{-4i\varphi_k}\tanh^2 r_k+1}
  {2(e^{-4i\varphi_k}\tanh^2 r_k-1)},
  \quad
  B\left(r_k,\varphi_k\right)=\frac{2 e^{-2i\varphi_k}\tanh r_k}{e^{-4 i \varphi_k}
      \tanh^2 r_k-1}.
\end{align}
Initially $r_k=0$, so $A=-1/2$ and $B=0$, and
$\Psi_{\bm k}(q_{\bm k},q_{-{\bm k}})\propto e^{-q_{\bm
    k}^2/2}e^{-q_{-{\bm k}}^2/2}$. Each mode ${\bm k}$ and $-{\bm k}$
is decoupled and placed in their ground state (namely, the
Bunch-Davies vacuum mentioned above). Then, the state evolves, $r_k$
becomes non-vanishing and $\Psi_{\bm k}(q_{\bm k},q_{-{\bm k}})$ can no longer
be written as a product $\Psi(q_{\bm k})\Psi(q_{-{\bm k}})$. This is
of course another manifestation of the fact that the state becomes
entangled.

The wave-functional $\Psi_\mathrm{2 sq}$ can also be written in the
basis $\vert \zeta_{\bm k}^{_\mathrm{R}},\zeta_{\bm
  k}^{_\mathrm{I}}\rangle $, where one defines $\hat{\zeta}_{\bm
  k}\equiv (\hat{\zeta}_{\bm k}^{_\mathrm{R}}+i\hat{\zeta}_{\bm
  k}^{_\mathrm{I}})/\sqrt{2}$, which implies that
\begin{align}
  \label{eq:linkqzeta}
  z\hat{\zeta}_{\bm k}^{_\mathrm{R}}=\frac{1}{\sqrt{2}}
  \left(\hat{q}_{\bm k}+\hat{q}_{-{\bm k}}\right), \quad 
z\hat{\zeta}_{\bm k}^{_\mathrm{I}}=\frac{1}{k\sqrt{2}}
  \left(\hat{\pi}_{\bm k}-\hat{\pi}_{-{\bm k}}\right).
\end{align}
In that case, $\Psi_\mathrm{2 sq}[\eta,\zeta]=\prod _{{\bm k}}
\Psi_{\bm k}(\zeta_{\bm k}^{_\mathrm{R}})\Psi_{\bm k}(\zeta_{\bm
  k}^{_\mathrm{I}})$, where the individual wave-functions can be
expressed as $\Psi_{\bm k}(\zeta_{\bm k}^s)\equiv \Psi_{\bm
  k}^s=N_{\bm k}e^{-\Omega _{\bm k}(a\zeta_{\bm k}^s)^2}$, where
$\vert N_{\bm{k}}\vert =\left(2\Rea \Omega_{\bm{k}}/\pi\right)^{1/4}$
and $s=\mathrm{R,I}$. The behavior of $\Omega_{\bm k}(\eta)$ is
determined by the Schr\"odinger equation, which leads to $\Omega_{\bm
  k}'=-2i\Omega_{\bm k}^2+i\omega^2(k,\eta)/2$, where we remind that
$\omega^2(k,\eta)$ is the time-dependent fundamental frequency of each
oscillator. Several remarks are in order at this point. First, the
wave-functional $\Psi_\mathrm{2 sq}[\eta,\zeta]$ can be obtained from
$\Psi_\mathrm{2 sq}[\eta,q]$ by canonical
transformation~\cite{Martin:2007bw,Grain:2019vnq}.
Second, finding the time dependence of
the function $\Omega_{\bm k}(\eta)$ is clearly equivalent to solving
the equation of motion~(\ref{eq:equv}). Third, given the previous
considerations about entanglement, it may seem surprising that $\Psi_{\bm
  k}(\zeta_{\bm k}^{_\mathrm{R}},\zeta_{\bm k}^{_\mathrm{I}})$ can be
written in a separable form, as a product of $\Psi_{\bm k}(\zeta_{\bm k}^{_\mathrm{R}})$
and $\Psi_{\bm k}(\zeta_{\bm k}^{_\mathrm{I}})$. But, in fact,
entanglement depends on how a system is divided into two bipartite
sub-systems. This is confirmed by a calculation of the quantum discord
which may be vanishing for a partition and non-vanishing for
another~\cite{Martin:2015qta}. Finally, in the wave-functional approach,
the two-point correlation function that was calculated in
\Eq{eq:twopointheisen} in the Heisenberg picture can be obtained with
the following formula
\begin{align}
  \label{eq:twopointshrodi}
  \left\langle 0\left \vert \zeta ^2\left(\eta,{\bm x}\right) \right
  \vert 0\right \rangle=
\int \prod_{\bm k} \dd \zeta_{\bm k}^{_\mathrm{R}}
\, \dd \zeta_{\bm k}^{_\mathrm{I}}\, \Psi_{\bm
  k}^*(\zeta_{\bm k}^{_\mathrm{R}},\zeta_{\bm k}^{_\mathrm{I}})
\, \zeta ^2\left(\eta,{\bm x}\right)
\, \Psi_{\bm
  k}(\zeta_{\bm k}^{_\mathrm{R}},\zeta_{\bm k}^{_\mathrm{I}}).
\end{align}
This leads to the power spectrum
\begin{align}
  \label{eq:spectrum:omega}
{\cal P}_{\zeta}(k)&=\frac{k^3}{2\pi^2}
\frac{1}{4\Rea \Omega_{\bm k}},
\end{align}
which can be checked to match the one obtained in \Eq{eq:twopointheisen}.

Having explained how the theory of quantum-mechanical inflationary
perturbations can be used to calculate the power spectrum ${\cal
  P}_\zeta(k)$ of the fluctuations, let us now briefly describe how
this power spectrum can be related to astrophysical observations. In
modern Cosmology, there exist many different observables that probe
various properties of the universe. Among the most important ones is
clearly the CMB temperature anisotropy mentioned before. It is the
earliest probe, that is to say the closest to the inflationary epoch,
that we have at our disposal. The CMB radiation is a relic thermal
radiation emitted in the early universe at a redshift of
$z_\mathrm{lss}\simeq 1100$. Since the early universe is extremely
homogeneous and isotropic, the temperature of this radiation (namely
$\sim 2.7$K) is almost independent of the direction towards which we
observe it. In fact, the early universe is not exactly homogeneous and
isotropic, precisely because of the presence of the curvature
perturbations discussed before. They manifest themselves by tiny
variations of the CMB temperature, at the level $\delta T/T\simeq
10^{-5}$. The CMB anisotropy is thus the earliest observational
evidence of curvature perturbations. More explicitly, the Sachs-Wolfe
effect~\cite{Sachs:1967er} relates the curvature perturbation
$\hat{\zeta}_{\bm k}$ to the temperature anisotropy $\widehat{\delta
  T}/T$ through the following formula
\begin{align}
\label{eq:sw}
\widehat{\frac{\delta T}{T}}({\bm e})
=& \int \frac{{\rm d}{\bm k}}{(2\pi)^{3/2}}
\left[F({\bm k})+i{\bm k}\cdot {\bm e}\, G({\bm k})\right]
\hat{\zeta}_{\bm k}(\eta_{\rm end})
e^{-i {\bm k}\cdot 
{\bm e}(\eta_{\rm lss}-\eta_0)
+i{\bm k}\cdot {\bm x}_{0}}\, ,
\end{align}
where ${\bm e}$ is a unit vector that indicates the direction on the
celestial sphere towards which the observation is performed. The
conformal times $\eta_{\rm lss}$ and $\eta_0$ are the last scattering
surface (lss) and present day ($0$) conformal times, respectively. The
vector ${\bm x}_0$ represents the Earth's location. The quantities $F({\bm
  k})$ and $G({\bm k})$ are the so-called form factors, which encode the
evolution of the perturbations after they have crossed in the Hubble
radius after inflation. In practice, the temperature anisotropy given by
Eq.~(\ref{eq:sw}) can be Fourier expanded in terms of the spherical
harmonics $Y_{\ell m}$, namely
\begin{align}
  \widehat{\frac{\delta T}{T}}({\bm e})
  =\sum _{\ell=2}^{+\infty}\sum _{\ell=-m}^{\ell=m}
  \hat{a}_{\ell m}Y_{\ell m}({\bm e}) .
  \end{align}
Using the
completeness of the spherical harmonics basis and Eq.~(\ref{eq:sw}), it
is easy to establish that, on large scales, namely in the limit
$F({\bm k})\rightarrow 1$ and $G({\bm k})\rightarrow 0$, one has
\begin{align}
\label{eq:alm}
  \hat{a}_{\ell m}=\frac{4\pi}{(2\pi)^{3/2}}e^{i\pi \ell/2}
  \int_{\setR^{3}} \dd {\bm k} \, \hat{\zeta}_{\bm k}(\eta_\mathrm{lss})\, j_\ell
       [k(\eta_\mathrm{lss}-\eta_0)]\, Y_{\ell m}^*({\bm k}),
       \end{align}
where $j_\ell$ is a spherical Bessel function. A CMB map is nothing
but a collection of numbers $a_{\ell m}$. The statistical properties of
a map is characterized by its powers spectrum, which can be written as
\begin{align}
  \left \langle 0\left \vert \widehat{\frac{\delta T}{T}}({\bm e}_1)
  \widehat{\frac{\delta T}{T}}({\bm e}_2)\right \vert 0\right \rangle
  =\sum_{\ell=2}^{+\infty}\frac{2\ell+1}{4\pi}
  C_\ell P_\ell\left(\cos \delta\right),
\end{align}
where $P_\ell$ is a Legendre polynomial and $\delta $ the angle
between the direction ${\bm e}_1$ and ${\bm e}_2$. The coefficients
$C_\ell$ are the so-called multipole moments and are related to the
$\hat{a}_{\ell m}$ by $\langle 0\vert \hat{a}_{\ell m}\hat{a}_{\ell '
  m'}^{\dagger}\vert 0 \rangle=C_\ell \delta_{\ell \ell'}\delta
_{mm'}$. From \Eq{eq:alm}, one can also write
\begin{align}
  \label{eq:multipole}
  C_\ell=\int _0^{+\infty}\frac{\dd k}{k}\, {\cal P}_\zeta(k)\, j_\ell^2
       [k(\eta_\mathrm{lss}-\eta_0)],
\end{align}
thus establishing the relation between the power spectrum ${\cal
  P}_\zeta$ and a CMB map. Let us emphasize again that this relation
is in fact oversimplified since it is obtained in the large-scale
limit. In order to be realistic, one should take into account the
behavior of the perturbations once they re-enter the Hubble radius
after inflation which, technically, implies to consider the full form
factors $F({\bm k})$ and $G({\bm k})$. This is a non-trivial task,
which requires numerical calculations. It leads to a modulation of the
signal and to the appearance of oscillations or peaks in the multipole
moments, the so-called Doppler or acoustic peaks.

\section{Motivations}
\label{sec:motivation}

The previous framework is usually viewed as very efficient. In
particular, the multipole moments~(\ref{eq:multipole}) calculated with
the inflationary power spectrum fit very well the CMB maps obtained by
the Planck satellite. Why, then, is the theory of quantum
perturbations still considered by some as unsatisfactory or
incomplete? The main reason is related to foundational issues in
Quantum Mechanics, more precisely to the so-called measurement
problem. In the context of inflation, this discussion is especially
subtle and, hence, interesting for the following reasons.

On one hand, the inflationary perturbations are placed in a Gaussian
state, which means that the corresponding Wigner function is also a
Gaussian and, therefore, is
positive-definite~\cite{2004JOptB...6..396K}. The Wigner function can
thus be used and interpreted as a classical stochastic
distribution~\cite{Polarski:1995jg, Albrecht:1992kf, Martin:2015qta},
in the sense that any two-point Hermitian correlation function can
always be reproduced with this Gaussian classical stochastic
distribution~\cite{Martin:2015qta}. This is also the case for any
higher-order correlation function involving position only, in
particular, any function of the curvature perturbation. It is
sometimes argued that these properties require large quantum squeezing
but, in fact, a large value of $r$ is needed only for those higher
correlation functions mixing position and momentum (which are, in any
case, not observable since they involve the momentum, that is to say
the decaying mode of the
perturbations~\cite{Martin:2015qta}). Nevertheless, the fact that all
observable correlation functions can be reproduced by stochastic
averages is often interpreted as the signature that a
quantum-to-classical transition has taken place.

On the other hand, we have argued before that the perturbations are
very ``quantum''. They are placed in a very strongly squeezed state,
which is a highly entangled state. Indeed, in the limit of infinite
squeezing, a squeezed state tends to an Einstein Podolski Rosen state,
which was used in the EPR argument to discuss the ``weird'' (namely
non-classical) features of Quantum Mechanics. It is hard to think
about a system that would be more ``quantum'' than this one!  As a
consequence, the statement that the system has become classical
should, at least, require some clarification. In fact, characterizing
the system as ``classical'' because some correlation functions can be
mimicked with a stochastic Gaussian process suffers from a number of
problems. First, even in the large-squeezing limit, there are
so-called ``improper operators'', for which the Weyl transform takes
some values outside the spectrum of the operator.  The measurement of
these operators can never be described with a classical stochastic
distribution~\cite{Revzen2006-REVTWF}. This, for instance, leads to
the possibility to violate Bell inequalities even if the Wigner
function always remains positive, a property which clearly signals
departure from
classicality~\cite{2005PhRvA..71b2103R,Martin:2016tbd,Martin:2017zxs}. In
fact, the question of whether Bell's inequality can be violated in a
situation where the Wigner function is positive-definite has been a
concern for a long time and was discussed by John Bell
himself~\cite{1986NYASA.480..263B}. The corresponding history, told in
\Refa{Martin:2019wta}, is a chapter of the history of Quantum Mechanics
and is associated to the difficulties to define a classical
limit. Second, there is the definite outcome question. With the theory
of decoherence~\cite{Zurek:1981xq,Schlosshauer:2003zy}, it is possible
to understand why we never observe a superposition of states
corresponding to macroscopic configurations but this is not sufficient
to explain why a specific state is singled out in the measurement
process. In some sense, with the help of quantum decoherence, the
quantum measurement problem has been reduced to the definite outcome
problem, which is at the core of the foundational issues of Quantum
Mechanics. In a cosmological context, let us mention that decoherence
has been studied and it has been suggested that it is likely to be at
play during
inflation~\cite{Burgess:2006jn,Martin:2018zbe,Martin:2018lin}. But the
definite outcome problem is still there and is neither solved by
decoherence (as already mentioned), nor by the emergence of
``classical'' stochastic properties as described above.

In fact, one could even argue that this question, in the context of
inflation and Cosmology, is worst than in the lab for the following
reasons. We have seen that the operators $\widehat{\delta T}/T({\bm
  e})$ (one for each direction ${\bm e}$) are observable
quantities. Since a measurement of these observables has been
performed by the COBE, WMAP and Planck satellites, according to the
basic postulates of Quantum Mechanics, the system must be placed in
one of the eigenstates of $\widehat{\delta T}/T({\bm e})$, that we
denote $\ketmapPlanck_{\rm Planck}({\bm e})$, and that satisfies
\begin{align}
\label{eq:eigencmb}
\widehat{\frac{\delta T}{T}}({\bm e})
\ketmapPlanck_{\rm Planck}({\bm e})=\frac{\delta T}{T}({\bm e})
\ketmapPlanck_{\rm Planck}({\bm e}).
\end{align}
However, the state $\vert \Psi_{2\, {\rm sq}}\rangle$ [recall that
  this state is defined in Eq.~(\ref{eq:quantumstate})] is not an
eigenstate of the temperature anisotropy operator. This can be
established with a direct and explicit calculation, but a physically
more intuitive method is based on the concept of
symmetry~\cite{Castagnino:2014cpa}. In order to simplify the
discussion, let us first use the fact that the curvature perturbation
can be viewed as a massless scalar field living in a
Friedmann-Lema\^itre-Robertson-Walker (FLRW) universe with an action
given by $S=-1/2\int \dd^4 {\bm x} \, \sqrt{-g} \, g^{\mu \nu}\,
\partial_{\mu }\zeta \, \partial_\nu \zeta$. Then, let us define the
$4$-momentum operator by
\begin{align}
  \hat{P}_{\mu}=-\int \dd ^3{\bm x}\, \sqrt{{}^{(3)}g}\, \hat{T}^0{}_{\mu},
  \end{align}
where $\hat{T}_{\mu \nu}$ is the stress energy tensor that can be
calculated from the action given above, $\hat{T}_{\mu
  \nu}=\partial_\mu \hat{\zeta} \partial_\nu \hat{\zeta}-g_{\mu
  \nu}g^{\alpha \beta}\partial_\alpha \hat{\zeta}
\partial_{\beta}\hat{\zeta}/2$ and ${}^{(3)}g$ the determinant of the
three-dimensional spatial metric. In cosmic time, one can check that
$\hat{P}_0$ exactly corresponds to the generator of the time evolution
of the system, namely the Hamiltonian. On the other hand, the
generator of the space translation along $x_i$ is given by
$\hat{P}_i=a\int \dd^3{\bm x}\, \dot{\hat{\zeta}}\,
\partial_i\hat{\zeta}$. Expressed in terms of creation and
annihilation operators, one obtains $\hat{P}_i\propto \int \dd {\bm k}
\, k_i \hat{c}_{\bm k}^{\dagger}\hat{c}_{\bm k}$. It follows
immediately from this expression that $\hat{P}_i\vert 0\rangle =0$ and
the same conclusion would be obtained by applying the generator of
rotations (angular momentum operator). This expresses the fact that
the vacuum state is homogeneous and isotropic, \ie it possesses the
symmetries of the FLRW background. Moreover, one has
$[\hat{H}_\mathrm{free},\hat{P}_i]=0$ and
$[\hat{H}_\mathrm{int},\hat{P}_i]=0$, hence $[\hat{H},\hat{P}_i]=0$,
which implies that the homogeneity and isotropy of the state is
preserved during cosmic expansion. As a result, one has
$\hat{P}_i\vert \Psi_{2\, {\rm sq}}\rangle=0$, and $\vert \Psi_{2\,
  {\rm sq}}\rangle$ still represents a universe without any
structure. Since $\hat{P}_i\ketmapPlanck_{\rm Planck}({\bm e})\neq 0$,
the transition between the two-mode squeezed
state~(\ref{eq:quantumstate}) and a state corresponding to a specific
outcome for CMB anisotropies, namely
\begin{align}
\label{eq:collapse}
\vert \Psi_{2\, {\rm sq}}\rangle=
\sum_{ \map}c( \map)  \ketmap \rightarrow 
\ketmapPlanck_{\rm Planck}({\bm e}),
\end{align}
cannot be generated by the Schr\"odinger equation. This is a
concrete manifestation of the measurement and single outcome problems
of Quantum Mechanics, which appear much more serious in a
cosmological context than in standard lab
situations, since the transition~(\ref{eq:collapse}) seems to have
taken place in the absence of any observer.

This leads to a first motivation for considering collapse models in
Cosmology. In this class of theories, the collapse of the wave-function
is a dynamical process controlled by a modified Schr\"odinger
equation, which does not rely on having an observer. 
%It is thus natural to apply them to the theory of quantum-mechanical perturbations produced during a phase of inflation. 
Another motivation is related to the fact that collapse models are
falsifiable. Indeed, since they are based on a modified Schr\"odinger
equation, they imply different predictions than standard Quantum
Mechanics. Given that the inflationary predictions can be accurately
tested with astrophysical data, one can then use them in order to test
Quantum Mechanics and collapse models in physical regimes that are
completely different from those usually probed in the lab.
This also shows that solving the quantum measurement problem can have
concrete implications for comparing the inflationary paradigm with the
data. Therefore, the question of how a particular realization is
produced is not of academic interest only, since it may also alter the
properties of the possible realizations themselves.

\section{Inflation and Collapse}
\label{sec:infcoll}

There is no unique collapse model but different versions that come in
different flavors. They are, however, all based on a modified
Schr\"odinger equation that, for a non-relativistic system,
reads~\cite{Ghirardi:1989cn}
\begin{align}
  \label{eq:csleq}
  \dd \Psi(t,{\bm x})&=\biggl[-i\hat{H}\dd t
    +\frac{\sqrt{\gamma}}{m_0}
    \sum_i \left(\hat{C}_i-\left\langle \Psi\left\vert \hat{C}_i
    \right\vert \Psi\right\rangle\right)
    \dd W_i(t)-\frac{\gamma}{2m_0^2}\sum_i\left(\hat{C}_i
    -\left\langle \Psi\left\vert
    \hat{C}_i\right\vert \Psi\right\rangle\right)^2\dd t
    \biggr]\Psi(t,{\bm x}),
  \end{align}
where $\hat{H}$ is the Hamiltonian of the system and $\hat{\bm C}$ a
collapse operator to be chosen (with three components denoted
$\hat{C}_i$, $i=x,y,z$). The parameter $\gamma $ is a new fundamental
constant the dimension of which depends on the choice of $\hat{\bm
  C}$, and $m_0$ is a reference mass usually taken to be the mass of a
nucleon. Finally, $\dd W_i(t)$ is a stochastic noise with
$\mathbb{E}[\dd W_i(t)\dd W_j(t')]=\delta _{ij}\delta(t-t')$ where
$\mathbb{E}[.]$ denotes the stochastic average. Notice that the above
equation is not sufficient to define the CSL model because we have not
yet specified what the collapse operator is.

Then, let us consider a field $\hat{\zeta}(t,{\bm x})$ and here, of
course, we have in mind curvature perturbation. Quantum mechanically,
it is described by a wave-functional $\Psi[\zeta({\bm x})]$ and we
need to know which form the general dynamical collapse equation~(\ref{eq:csleq})
takes in this case. A first question that immediately arises is that
the above equation~(\ref{eq:csleq}) is, in principle, valid in the
non-relativistic regime only while one needs to go beyond since we
want to apply collapse models to Cosmology and Field Theory. Attempts
to develop a relativistic version of the collapse models are being
carried out, see \eg
\Refs{Ghirardi:1989cn,Tumulka:2005ki,Bedingham:2010hz,2014JSP...154..623B}
but they are not completed yet. Therefore, either one stops at this
stage and waits for a fully satisfactory relativistic version to come,
or one proceeds using reasonable assumptions, at the price of being
maybe on shaky grounds. Here, we use collapse theories in Cosmology
where there is a natural notion of time (the Hubble
flow). Technically, this often means that the relativistic equations
describing a phenomenon are well-approximated by the corresponding
non-relativistic equations only modified by the appearance of the
scale factor at some places. The prototypical example of such an
approach is ``Newtonian Cosmology'' for which the laws that describe
the time evolution of an expanding homogeneous and isotropic universe
can be deduced from Newtonian dynamics and gravitation. Although the
derivation is not strictly self-consistent it nevertheless provides
some intuitive insights and represents a valuable first step. In some
sense, here, we follow the same logic and, therefore, we will simply
postulate that Eq.~(\ref{eq:csleq}) can also be used in this context
where the Hamiltonian of the system is simply the
Hamiltonian~(\ref{eq:hamilton}) that is obtained from the theory of
relativistic cosmological perturbations.

In order to see what this implies in practice, it is convenient to
view space-like sections as an infinite grid of discrete points. In
this case, the functional can be interpreted as an ordinary function
of an infinite number of variables $v_i$,
$\Psi(\cdots,v_i,v_j,\cdots)$, where $v_i\equiv v({\bm x}_i)$ is the
value of the field at each point of the grid. Therefore, instead of
dealing with a three-dimensional index $i$ as before, we now deal with
an infinite-dimensional one. As a consequence, we can write an
equation similar to Eq.~(\ref{eq:csleq}) for $\Psi(v_i)$ where, now,
the operators $\hat{H}$ and $\hat{\bm C}$ are functions of the
``position'' $\hat{v}_i$ and ``momentum'' $\hat{p}_i=-i\partial
/\partial v_i$. Then, taking the continuous limit, ``$\sum_i
\rightarrow \int \dd {\bm x}_\mathrm{p}$'', we arrive at
\begin{align}
\label{eq:cslphys}
\kern-0.2em\dd \kern-0.2em\left\vert \Psi[\zeta({\bm x}_\mathrm{p})]
\right\rangle  \kern-0.2em & = 
 \kern-0.2em \biggl\lbrace \kern-0.2em -i \hat{H} \dd t + \kern-0.2em
\frac{\sqrt{\gamma}}{m_0} 
\displaystyle{\int}\kern-0.2em 
\dd \bm{x}_\mathrm{p}  \kern-0.1em\left[\hat{C} \kern-0.2em
\left(\bm{x}_\mathrm{p}\right)  \kern-0.2em
-  \kern-0.2em\left\langle \hat{C} \kern-0.2em
\left(\bm{x}_\mathrm{p}\right) 
\right\rangle \right]
\dd W_t \kern-0.2em\left(\bm{x}_\mathrm{p}\right)
-\frac{\gamma}{2m_0^2}\displaystyle{\int}\dd \bm{x}_\mathrm{p}
\left[\hat{C}
\left(\bm{x}_\mathrm{p}\right) 
- \left\langle \hat{C}
\kern-0.2em\left(\bm{x}_\mathrm{p}\right) 
\right\rangle \right]^2\dd t
\biggr\rbrace \left\vert \Psi[\zeta({\bm x}_\mathrm{p})]\right\rangle .
\end{align}
The quantity $\dd W_t({\bm x}_\mathrm{p})$ is still a stochastic noise
but we now have one for each point in space. A fundamental aspect of
the theory is to specify this noise, and each possibility corresponds
to a different version of the theory. A priori, as already mentioned,
the noise can be white or colored but, so far in the context of
Cosmology, only white noises have been considered. They satisfy
$\mathbb{E}[\dd W_t({\bm x}_\mathrm{p})\dd W_{t'}({\bm
    x}_\mathrm{p}')]=\delta({\bm x}_\mathrm{p}-{\bm
  x}_\mathrm{p}')\delta(t-t')$. Let us also notice that
$\bm{x}_\mathrm{p}$ denotes the physical coordinate, as opposed to the
comoving one ${\bm x}$ (${\bm x}_\mathrm{p}=a{\bm x}$) usually
employed in Cosmology, and in terms of which \Eq{eq:cslphys} takes the
form~\cite{Martin:2019jye}
\begin{align}
\label{eq:cslphysco}
\kern-0.2em\dd \kern-0.2em\left\vert \Psi[\zeta({\bm x})]\right\rangle
\kern-0.2em & = 
 \kern-0.2em \biggl\lbrace \kern-0.2em -i \hat{H} \dd t + \kern-0.2em
\frac{1}{m_0}\sqrt{\frac{\gamma}{a^3}} 
\displaystyle{\int}\kern-0.2em 
\dd \bm{x}\, a^3  \kern-0.1em\left[\hat{C} \kern-0.2em
\left(\bm{x}\right)  \kern-0.2em
-  \kern-0.2em\left\langle \hat{C} \kern-0.2em
\left(\bm{x}\right) 
\right\rangle \right]
\dd W_t \kern-0.2em\left(\bm{x}\right)
-\frac{\gamma}{2m_0^2}\displaystyle{\int}\dd {\bm x}\, a^3
\left[\hat{C}
\left(\bm{x}\right) 
- \left\langle \hat{C}
\kern-0.2em\left(\bm{x}\right) 
\right\rangle \right]^2\dd t
\biggr\rbrace \left\vert \Psi[\zeta({\bm x})]\right\rangle ,
\end{align}
where $\dd W_t({\bm x}_\mathrm{p})=a^{-3/2}\dd W_t({\bm x})$ so that
$\dd W_t({\bm x})$ is still white, namely $\mathbb{E}\left[\dd
  W_t({\bm x}) \dd W_{t'}({\bm x}')\right] =\delta ({\bm x}-{\bm
  x}')\delta (t-t')\dd t^2$. We emphasize that the above stochastic
equation is the usual CSL equation: it is just written down in a
situation where the number of variables becomes infinite.

Of course, we are not forced to describe the field $\hat{\zeta}({\bm
  x})$ in real space and we can also write it in Fourier space. In
that case, the wave-functional becomes a function of all Fourier
components of the field, $\Psi(\cdots, \zeta_{\bm k}, \zeta_{{\bm
    k}'}, \cdots)$, that is to say we deal, again, with the same
situation as described by Eq.~(\ref{eq:csleq}) but, now, with a
continuous index ${\bm k}$ instead of $i=x,y,z$. The advantage of this
approach is that, because we work in the framework of linear
perturbations theory, one can write the wave-function as $\Psi(\cdots,
\zeta_{\bm k}, \zeta_{{\bm k}'}, \cdots) =\prod_{\bm k}\Psi_{\bm
  k}^\mathrm{R}\Psi_{\bm k}^\mathrm{I}$. As explained before, we have
used the notation $s=\mathrm{R},\mathrm{I}$ so that $\Psi_{{\bm
    k}}^{s}\equiv \Psi(\zeta_{\bm k}^s)$. This is the great advantage
of going to Fourier space compared to real space: it drastically
simplifies the wave-function. One may, however, wonder whether the
non-linearities necessarily present in the theory (recall that the new
terms in the Schr\"odinger equation are necessarily stochastic and
non-linear) could bring to naught the technical convenience of using
the Fourier transform. Usually, only when a theory is linear, the
Fourier modes evolve independently (no mode coupling) and it is useful
to go to Fourier space. This corresponds to a situation where the
Hamiltonian is quadratic. A point, which is usually not very well
appreciated, is that this does not necessarily imply the absence of
interactions. It is true that, in field theory, interactions are
associated with non-quadratic terms in the action but one exception is
the interaction of a quantum field with a classical source. In this
case, the action remains quadratic but the fundamental frequency of
the system acquires a time dependence given by the source. This is
typically the case for the Schwinger
effect~\cite{Schwinger:1951nm,Martin:2007bw} but also for
Cosmology. In this last situation, the source is just the dynamics of
the background space-time itself. In the following, we restrict
ourselves to quadratic Hamiltonians since this is sufficient to
describe cosmological perturbations during inflation (of course, if
one wants to calculate higher-order statistics, such as
Non-Gaussianities, then non-linear terms in the Hamiltonian must be
taken into account).

However, in the present situation, even if one restricts oneself to
quadratic Hamiltonians, one also has the extra non-linear and
stochastic terms in the modified Schr\"odinger equation and, as
noticed above, there is the concern that they could be responsible for
the appearance of mode couplings. Fortunately, this is not necessarily the
case. Indeed, if one recalls that the Hamiltonian of the system reads
$\hat{H}=\int _{\setR^{3+}}\dd {\bm k} \sum_{s=\mathrm{R},\mathrm{I}}
\hat{H}_{{\bm k}}^s$ and if one introduces the Fourier transform of
the collapse operator, $\hat{C}({\bm x})=(2\pi)^{-3/2} \int \dd {\bm
  k} \, \hat{C}({\bm k})e^{-i {\bm k}\cdot {\bm x}}$ (and a similar
formula for the noise), then straightforward calculations lead
to~\cite{Martin:2019jye}
\begin{align}
\label{eq:cslphysfourier}
 \kern-0.2em\dd \kern-0.2em\left\vert \Psi\right\rangle  \kern-0.2em & = 
 \int _{\setR^{3+}}\dd {\bm k} \sum_{s=\mathrm{R},\mathrm{I}}
 \kern-0.2em \biggl\lbrace \kern-0.2em -i \hat{H}_{{\bm k}}^s \dd t + \kern-0.2em
\frac{\sqrt{\gamma a^3}}{m_0} 
\left[\hat{C}^s \kern-0.2em
\left(\bm{k}\right)  \kern-0.2em
-  \kern-0.2em\left\langle \hat{C}^s \kern-0.2em
\left(\bm{k}\right) 
\right\rangle \right]
\dd W_t^s \kern-0.2em\left(\bm{k}\right)
-\frac{\gamma a^3}{2m_0^2}
\left[\hat{C}^s
\left(\bm{k}\right) 
- \left\langle \hat{C}^s
\kern-0.2em\left(\bm{k}\right) 
\right\rangle \right]^2\dd t
\biggr\rbrace \left\vert \Psi_{{\bm k}}^s\right\rangle .
\end{align}
We see that, if the Fourier transform of the collapse operator, $\hat{C}^s
\left(\bm{k}\right) $, only contains operators acting in the $\bm{k}$ subspace (this is notably the case if $C(\bm{x})$ is a linear combination of the phase-space variables), then we can write a CSL equation for each Fourier mode. In
other words, it seems that the presence of the extra stochastic and
non-linear terms does not necessarily destroy the property that the modes still
evolve separately~\cite{Martin:2019jye}. In order to better understand
the origin of this property, let us come back to
Eq.~(\ref{eq:csleq}). Let us assume that we are in the particular
situation where $\hat{H}=H(\hat{\bm x},\hat{\bm
  p})=H_1(\hat{x}_1,\hat{p}_1)+H_2(\hat{x}_2,\hat{p}_2)
+H_3(\hat{x}_3,\hat{p}_3)$ and $\hat{C}_i=C_i(\hat{\bm x},\hat{\bm
  p})=C_i(\hat{x}_i,\hat{p}_i)$, namely the component $\hat{C}_i$ only
depends on $\hat{x}_i$ and $\hat{p}_i$ [in other words, we do not
  have, for instance, $\hat{C}_x=C_x(\hat{y},\hat{p}_y)$]. Then
writing $\Psi=\prod_i \Psi_i(x_i)$, it is easy to show that
\begin{align}
  \label{eq:csleachdim}
  \dd \Psi_i&=\left[-i\hat{H}_i\dd t +\frac{\sqrt{\gamma}}{m_0}
    \left(\hat{C}_i-\left\langle \Psi_i\left\vert \hat{C}_i
    \right\vert \Psi_i\right\rangle\right) \dd
    W_i-\frac{\gamma}{2m_0^2}\left(\hat{C}_i-\left\langle
    \Psi_i\left\vert \hat{C}_i\right\vert
    \Psi_i\right\rangle\right)^2\dd t \right]\Psi_i,
  \end{align}
where we have used the fact that
\begin{align}
  \label{eq:meancollapse}
  \left\langle \Psi \left\vert \hat{C}_i\right\vert \Psi \right\rangle
  &=\left\langle \prod_j \Psi_j\left\vert
  \hat{C}_i\right\vert \prod_k \Psi_k\right \rangle
=\left\langle \prod_{j\neq i} \Psi_j\biggl\vert
  \prod_{k\neq i} \Psi_k\right \rangle
\left\langle \Psi_i \left\vert \hat{C}_i\right\vert \Psi_i \right \rangle
=\left\langle \Psi_i \left\vert \hat{C}_i\right\vert \Psi_i \right \rangle.
\end{align}
We see that we can write an independent equation for each $\Psi_i$. In
inflationary perturbations theory, if the collapse operator is a linear combination of the field phase-space variables, the two properties needed to obtain
this independent equation are also satisfied, namely the Hamiltonian
is a sum of the Hamiltonians for each Fourier mode and $\hat{C}^s({\bm
  k})$ only depends on ${\bm k}$ and not on other modes. This is the
reason why one can obtain an equation~(\ref{eq:cslphysfourier}) for
each Fourier mode.

Then comes the choice of the collapse operator $\hat{C}({\bm
  x}_\mathrm{p})$. Many different possibilities have been discussed in
the literature and each of them correspond to a different version of
the theory. In the context of standard Quantum Mechanics, if
$\hat{C}({\bm x}_\mathrm{p})$ is the position operator, then we have
Quantum Mechanics with Universal Position Localization (QMUPL) while
if $\hat{C}({\bm x}_\mathrm{p})$ is the mass density operator, we deal with the
Continuous Spontaneous Localization (CSL)
model~\cite{Ghirardi:1989cn}. In the context of Field Theory and
Cosmology, two choices have been studied. The first one corresponds to
$\hat{C}^s({\bm k})\propto a^p \hat{\zeta}_{\bm k}^s$, where $p$ is a free
parameter. Since, in some sense, field amplitude plays the role of
position, this case represents the field-theoretic version of
QMUPL. Except for $p$, this version is characterized by one parameter,
$\gamma $. The other possibility is CSL, which relies on coarse-graining
the mass density over the distance $r_\mathrm{c}$. This corresponds to
\begin{align}
  \hat{C}({\bm x})=\left(\frac{a}{r_\mathrm{c}}\right)^3
  \frac{1}{(2\pi)^{3/2}}\int \dd {\bm y} \, \hat{\delta}_\mathrm{g}({\bm x}+{\bm y})
  e^{-\frac{\vert{\bm y}\vert^2a^2}{2r_\mathrm{c}^2}},
    \end{align}
where $\hat{\delta}_\mathrm{g}$ is the energy density contrast
relative to a ``Newtonian'' time slicing (see the beginning of the
next section for a more complete discussion). At this point, we meet
again the problem that a fully relativistic and covariant collapse
model is not available. Indeed, the definition of energy density is
not unique in General Relativity and an infinite number of other
choices could have been contemplated, by considering the energy
density contrast relative to other
slicings~\cite{Martin:2019jye}. Without additional criterions, there
is presently no mean to decide which version makes more
sense. However, what can be done is to constrain these different
versions with CMB data. In fact, and we come back to this question in
the next section, Sec.~\ref{sec:comparison}, we can show that the
situation is not as problematic as it may seem and that (almost) all
possible choices lead to the same result. In this sense, the results
obtained in the following are rather generic.

Once the collapse operator and the noise have been chosen,
\Eq{eq:cslphysfourier} is entirely specified and the next step is then
to solve it. The solution is given by a wave-function evolving
stochastically in Hilbert space. As discussed above, the initial
conditions are Gaussian and the Hamiltonian being quadratic, the
Gaussian character of the wave-function is preserved in
time. Therefore, without loss of generality, one can write the most
general stochastic wave-function as
\begin{align}
\label{eq:stochawf}
\Psi_{\bm{k}}^s\left(\zeta_{\bm{k}}^s\right)&= \vert
N_{\bm{k}}\left(\eta\right)\vert \exp\Bigl\lbrace -\Rea
\Omega_{\bm{k}}\left(\eta\right)z^2
\left[\zeta_{\bm{k}}^s-\bar{\zeta}_{\bm{k}}^s\left(\eta\right)\right]^2
+i\sigma_{\bm k}^s(\eta)+iz\chi_{\bm k}^s(\eta) \zeta_{\bm{k}}^s -iz^2\Ima
\Omega_{\bm k}(\eta ) \left(\zeta_{\bm{k}}^s\right)^2\Bigr\rbrace,
\end{align}
where the free functions $\Omega_{\bm k}(\eta)$,
$\bar{\zeta}_{\bm{k}}^s(\eta)$, $\sigma_{\bm k}^s(\eta)$ and
$\chi_{\bm k}^s(\eta)$ are (a priori) stochastic quantities.

Let us now discuss how collapse models can be, in the context of
Cosmology, related to observations. This needs to be carefully studied
since we now have two ways to calculate averages, the quantum average
and the stochastic average. For instance, the quantum average of a
given observable ${\cal O}(\hat{\zeta}_{\bm k}^s)$, $\langle {\cal
  O}(\hat{\zeta}_{\bm k}^s)\rangle\equiv \int \vert \Psi_{\bm k}^s
\vert^2 {\cal O}(\zeta _{\bm k}^s) \dd \zeta_{\bm k}^s$, which, in the
standard context, would be a number is, here, a stochastic
quantity. So only $\mathbb{E}[\langle{\cal O}(\hat{\zeta}_{\bm
    k}^s)\rangle]=\int \mathbb{E}[\vert \Psi_{\bm k}^s \vert^2 ]{\cal
  O}(\zeta _{\bm k}^s) \dd \zeta_{\bm k}^s$ is a number. The quantity
\begin{align}
  \vert \Psi_{\bm k}^s(\zeta_{\bm k}^s)\vert^2=z\sqrt{\frac{2 \Rea \Omega_{\bm k}}{\pi}}
  \exp\left[-2z^2 \Rea \Omega_{\bm k}
    \left(\zeta_{\bm k}^s-\bar{\zeta}_{\bm k}^s\right)^2\right],
  \end{align}
which is centered at $\bar{\zeta}_{\bm k}^s$ and has width $(4 z^2
\Rea \Omega_{\bm k})^{-1}$, describes a Gaussian wave-packet whose
mean and variance evolve stochastically (in fact, in the particular
case considered here, it turns out that the variance is a
deterministic quantity and that only the mean is
stochastic). Therefore, for a specific realization, one expects, as
time passes, that $\vert \Psi_{\bm k}^s(\zeta_{\bm k}^s)\vert^2$
stochastically shifts its position $\bar{\zeta}_{\bm k}^s(\eta)$ while
its width decreases until $\bar{\zeta}_{\bm k}^s$ settles down to a
particular position $\bar{\zeta}_{\bm k}^s(\eta_\mathrm{coll})$, with
an (almost) vanishing width. In this way, the macro-objectification
problem of Quantum Mechanics is solved and a single outcome has been
produced. The interest of this approach for Cosmology is that it does
so without invoking the presence of an observer, and only thanks to
the modified dynamics of the wave-function. If one then considers
another realization, a qualitatively similar behavior is observed but,
of course, the final value $\bar{\zeta}_{\bm k}^s(\eta_\mathrm{coll})$
(in fact the whole trajectory) needs not be the same. If we repeat
many times the same experiment and have at our disposal many
realizations, one can then calculate, say, $\mathbb{E}[\langle
  \hat{\zeta}_{\bm k}^s\rangle]=\mathbb{E}[\bar{\zeta}_{\bm k}^s]$ or
$\mathbb{E}[\langle \hat{\zeta}_{\bm{k}}^s \rangle^2] =
\mathbb{E}[\bar{\zeta}_{\bm k}^s{}^2]$. This allows us to calculate
the dispersion of $\bar{\zeta}_{\bm k}^s$ according to
\begin{align}
\label{eq:Pv:def}
{\cal P}_{\zeta}(k) =\frac{k^3}{2\pi^2}
\left\{\mathbb{E} \left[{{\bar{\zeta}_{\bm{k}}^s}}{}^2\right]
-\mathbb{E}^2\left[{{\bar{\zeta}_{\bm{k}}^s}}\right]\right\},
\end{align}
which makes the connection with the previous considerations.

In fact, in Cosmology, a legitimate question is why the above-defined
dispersion ${\cal P}_\zeta$ is equivalent to (or, even, has something
to do with) the power spectrum of curvature perturbations. Indeed, in
order to give an operational meaning to the above quantity, one needs
to have access to a large number of realizations. This is necessary if
one wants to identify the mathematical object $\mathbb{E}[.]$ with the
relative frequency of occurrence. Clearly, in Cosmology, we deal with
only one realization (one universe) and there is no way to repeat the
experiment. In fact, this question is by no mean an issue only for the
collapse models since, even in the standard approach, the predictions
are expressed in terms of ensemble averages.

Here, the key idea, admittedly not always explicitly stated in the
inflationary literature, is the use of an ergodic-like principle,
which consists in identifying ensemble averages with spatial
averages~\cite{Grishchuk:1997pk}. A very schematic description of this
procedure is as follows. For a given Fourier mode $\bm k$, one can
divide the celestial sphere into different patches, and construct an
estimate of the amplitude of the curvature perturbation at this
Fourier mode in each patch. Interpreting each patch as a different
realization, one can then calculate the ensemble average of these
``measurements'', which is thus nothing but a spatial average. In this
sense, ``repeating the experiment'' is replaced with ``looking at
different regions on the sky''. Obviously, to be able to evaluate the
Fourier mode $\bm k$ in a certain patch, the size of the patch has to
be larger than the wavelength associated to $\bm k$. However, the
celestial sphere being compact, only a finite number of patches with a
certain minimum size can be drawn on it. This is why the ensemble
average can be calculated only over a finite number of
``realizations'', and the larger the wavelength (\ie the smaller $k$)
is, the larger the patches need to be, hence the fewer
``realizations'' are available. This introduces an unavoidable error
which is called the ``cosmic variance'' in the Cosmology literature,
see \Refa{Grishchuk:1997pk} for more details.

\section{Comparison with Observations}
\label{sec:comparison}

\begin{figure}[t]
\begin{center}
\includegraphics[width=0.49\textwidth]{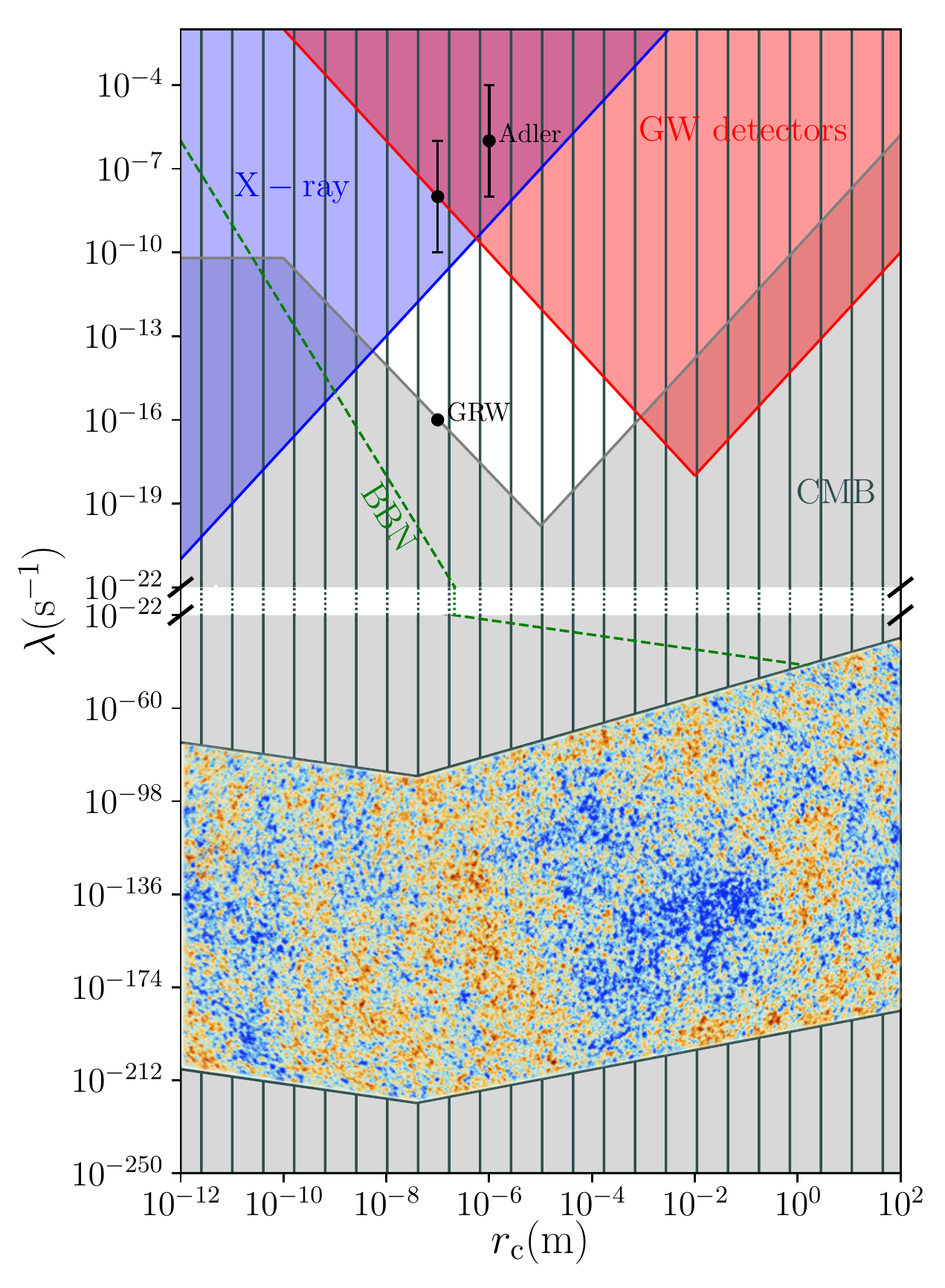}
\caption{Observational constraints on the two parameters $r_\uc$ and
  $\lambda$ of the CSL model obtained in \Refa{Martin:2019jye}. The
  white region is allowed by laboratory experiments while the ``CMB
  map'' region is allowed by CMB measurements. The green dashed line
  stands for the upper bound on $\lambda$ if inflation proceeds at the
  Big-Bang Nucleosynthesis (BBN) scale.}
\label{fig:mapCSL}
\end{center}
\end{figure}

In this section, we briefly discuss the observational status of
collapse models in Cosmology. As already mentioned, only few cases
have been investigated so far: QMUPL and CSL, both with a white noise
and using a naive generalization of non-relativistic collapse models
to field theory. A discussion of QMUPL in Cosmology can be found in
\Refs{Martin:2012pea,Das:2013qwa} and, here, we focus on CSL since
this is the model that has drawn the most
attention~\cite{Martin:2019jye}.

The CSL theory consists in assuming that the collapse operator is mass
or energy density. In a cosmological context, as already briefly
mentioned in the previous section, this corresponds to
$\hat{C}=\rho+\widehat{\delta \rho}$, where $\rho$ is the energy
density stored in the inflaton field and $\hat{\delta} \equiv
\widehat{\delta \rho}/\rho$ is the density contrast. In fact, only the
density contrast will be playing a role in what follows because, in
inflationary perturbations theory, $\rho$ is a classical quantity and,
therefore, cancels out in the modified Schr\"odinger equation. In
General Relativity, however, as already mentioned, there is no unique
definition for $\delta $. Nevertheless, see
Ref.~\cite{Martin:2019jye}, what matters is in fact the scale
dependence of $\delta $, in particular its behavior on large
scales. Conveniently, one can show that, for all reasonable choices,
all the $\delta $'s behave similarly (namely, in the same way as the
Newtonian density contrast ``$\delta _\mathrm{g}$'') except for one
particular case, the so-called ``$\delta_\mathrm{m}$'' density
contrast. Therefore, even if the choice of $\delta$ is ambiguous, the
final result turns out to be (almost) independent of this choice.

Once the collapse operator has been chosen, one can solve the modified
Schr\"odinger equation and calculate the CSL inflationary power
spectrum along the lines explained in the previous sections. This
power spectrum depends on the two CSL parameters $\gamma $ and
$r_\uc$. Quite intuitively, one finds that the extra CSL terms operate
only if the physical wavelength of a Fourier mode is larger than the
localization scale $r_\uc$. In an expanding universe, physical
wavelengths increase with time, so this implies that for any given
wavenumber $\bm k$, there is a time before which its physical
wavelength is smaller than $r_\uc$, hence the CSL corrections are
absent. This is a crucial feature since it guarantees that the usual
way of setting initial conditions in the Bunch-Davies vacuum, which is
a very important aspect of the inflationary paradigm, is still
available.

When the physical wavelength of a Fourier mode becomes larger than
$r_\uc$, the CSL terms become important and collapse occurs. This
generates the power spectrum~\cite{Martin:2019jye}
\begin{align}
  \label{eq:cslspectrum}
{\cal P}_{\zeta}(k)&=\frac{k^3}{2\pi^2} \frac{1}{4\Rea \Omega_{\bm
    k}\vert_{\gamma=0}} \biggl[1 +{\cal O}(1)\frac{\gamma}{m_0^2}
  \rho\, \epsilon_1 \left(\frac{r_\uc}{\ell_{_{\rm H}}}\right)_{\rm
    end}^\mathfrak{a} \left(\frac{k}{aH}\right)^\mathfrak{b}_\uend
  -\frac{\Rea \Omega_{\bm k}\vert_{\gamma=0}}{\Rea \Omega_{\bm k}}
  \biggr].
\end{align}
In the limit where $\gamma=0$, one checks that the power spectrum
vanishes, since no perturbation is being produced, in agreement with
the discussion presented in \Sec{sec:infcoll}. Let us also recall that
the ``standard'' result, obtained in the Copenhagen interpretation, is
given by \Eq{eq:spectrum:omega}, which matches the prefactor in
\Eq{eq:cslspectrum}, and that $\Rea \Omega_{\bm k}$ is proportional to
the inverse variance of the wave-packet. If $\gamma$ is sufficiently
large so that the collapse occurs, the width of the wave-function is
much smaller than what it would be in the unmodified theory, hence the
third term in the square brackets of \Eq{eq:cslspectrum} can be
neglected when compared to the first term. In that case, the power
spectrum takes the form of the standard result, plus a correction
proportional to $\gamma $. This CSL correction is also proportional to
$\rho \epsilon_1 $, where $\epsilon_1$ is the first slow-roll
parameter and $\rho$ the energy density at the end of inflation. Let
us recall that, during inflation, $\rho$ is quasi constant and can be
as large as
\begin{align}
  \rho \sim 10^{80} \mbox{g}\times \mbox{cm}^{-3}.
  \end{align}
We see here why Cosmology is a natural place to probe collapse
theories: it tests them in regimes that are completely different, in
terms of energy, time or length scales, than those relevant in the
lab. Since the amplitude of the CSL new terms are controlled by the
energy density, it makes sense to constrain them in physical
conditions where $\rho$ is as large as possible. This is why, for
instance, the CSL mechanism was also applied to neutron stars in
\Refa{Tilloy:2019nnp}. Primordial Cosmology is a situation where $\rho$
is even larger and, therefore, one can expect it to be even more
appropriate when it comes to establishing constraints on CSL.

The second crucial piece of information that comes from
\Eq{eq:cslspectrum} is that the CSL corrections are not scale
invariant. Their scale dependence is $\propto k^\mathfrak{b}$ where
$\mathfrak{b}=-1$ if the scale $r_\uc$ is crossed out during inflation
and $\mathfrak{b}=-10$ if $r_\uc$ is crossed out during the subsequent
radiation dominated era. In this last case, there is an additional
factor $\propto (r_\uc/\ell_{_\mathrm{H}})^\mathfrak{a}$, where
$\ell_{_\mathrm{H}}$ is the Hubble radius at the end of inflation,
with $\mathfrak{a}=-9$ (if $r_\uc$ is crossed out during inflation,
this term is not present and $\mathfrak{a}=0$). In other words,
detectable CSL corrections would be strongly incompatible with CMB
measurements. Since we have seen that they are typically very large,
we expect the constraints that can be inferred from them to be very
efficient.

These constraints are represented in \Fig{fig:mapCSL} in the space
$(r_\uc,\lambda)$ where $\lambda=\gamma/(8\pi^{3/2}r_\uc^3)$. In this
plot, the white region corresponds to the parameter space allowed by
lab experiments while the ``CMB map'' region corresponds to parameter
space allowed by CMB measurements. Evidently, the most striking
feature of the plot is that the two regions do not overlap. Taken at
face value, this implies that CSL is ruled out! However, this
conclusion should be toned down. First, we should notice that if the
collapse operator is taken to be $\delta_\mathrm{m}$, then the CMB
constraints are no longer in contradiction with the lab ones. Of
course, in some sense, $\delta_\mathrm{m}$ is ``of measure zero'' in
the space of density contrasts but, nevertheless, this shows that one
can find collapse operators for which CSL is rescued. Second, one has
to remember that we used a naive (too naive?) method to implement the
collapse mechanism in field theory. It could be that, when a truly
covariant version of collapse models is
available~\cite{Ghirardi:1989cn,Tumulka:2005ki,Bedingham:2010hz,2014JSP...154..623B},
the final result will be modified. For instance, the constraints on
the CSL parameters coming from the CMB constraints on one hand, and
from lab experiments on the other hand, operate at very different
energy scales. One could imagine that, in a field-theoretic context,
the CSL parameters run with the energy scale at which the experiment
is being performed, and that one cannot simply compare the constraints
obtained at different energies. Finally, we used a white noise in the
modified Schr\"odinger equation and it remains to be seen if using a
colored noise can modify the constraints obtained in
\Fig{fig:mapCSL}. For all these reasons, it is necessary to be
cautious and testing the robustness of the conclusions obtained here
will certainly be a major goal in the future.

\section{Conclusions}
\label{sec:conclusions}

Interestingly enough, collapse models advocated by Giancarlo Ghirardi
(and others) and cosmic inflation have almost the same age. Roughly
speaking, they were both introduced at the end of the seventies and
beginning of the eighties. Nevertheless, until recently, they had
never met. In this article, we have described the recent attempts to
apply collapse models to inflation. We have argued that there is a
good scientific case motivating those attempts. In particular, for
collapse models to be interesting and to insure proper localization,
the collapse operators must be related to the energy density. As a
consequence, the most efficient tests of collapse models will be in
physical situations where the energy density is as large as
possible. Without any doubt, this is to be found in the early
universe. We have shown that, indeed, the high-accuracy data now at
our disposal leads to extremely competitive constraints, that anyone
interested in collapse theories can no longer ignore. We hope this
will cause further investigations to test the robustness of these
results.

Finally, after $40$ years, collapse theories and cosmic inflation have
met and we are convinced that Giancarlo Ghirardi would have been
fascinated by the fact that his great insights about Quantum Mechanics
can even find applications in Cosmology.

\bibliography{CSL}
\end{document}

%% file: newcommands.tex
\newcommand{\ie}{{i.e.~}}

\newcommand{\eg}{\textsl{e.g.~}}

\newcommand{\dd}{\mathrm{d}}
\newcommand{\ee}{e}

\newcommand{\sss}[1]{{\scriptscriptstyle{#1}}}

\newcommand{\uend}{\mathrm{end}}
\newcommand{\uini}{\mathrm{ini}}

\newcommand{\uc}{\mathrm{c}}

\newcommand{\uS}{\mathrm{S}}

\newcommand{\usssS}{\sss{\uS}}

\newcommand{\nS}{n_\usssS}

\newcommand{\setR}{\mathbb{R}}

\newcommand{\Rea}{\Re \mathrm{e}\,}
\newcommand{\Ima}{\Im \mathrm{m}\,}

\newcommand{\GeV}{\mathrm{GeV}}

\newcommand{\beq}{\begin{equation}}
\newcommand{\eeq}{\end{equation}}
\newcommand{\bea}{\begin{eqnarray}}
\newcommand{\eea}{\end{eqnarray}}

\newlength{\wsingfig}
\newlength{\wdblefig}
\newlength{\wquadfig}
\newlength{\wtriplefig}
\newcommand{\Eq}[1]{Eq.~(\ref{#1})}
\newcommand{\Eqs}[1]{Eqs.~(\ref{#1})}
\newcommand{\Fig}[1]{Fig.~{\ref{#1}}}

\newcommand{\Refa}[1]{Ref.~{\cite{#1}}}
\newcommand{\Refs}[1]{Refs.~{\cite{#1}}}
\newcommand{\Sec}[1]{Sec.~\ref{#1}}